# Anthropomorphic User Interface Feedback in a Sewing Context and Affordances

Dr Pietro Murano
Computing, Science and Engineering
University of Salford
Salford, Gt Manchester, UK, M5 4WT

Tanvi Sethi
Computing, Science and Engineering
University of Salford
Salford, Gt Manchester, UK, M5 4WT

*Abstract*— The aim of the authors' research is to gain better insights into the effectiveness and user satisfaction of anthropomorphism at the user interface. Therefore, this paper presents a between users experiment and the results in the context of anthropomorphism at the user interface and the giving of instruction for learning sewing stitches. Two experimental conditions were used, where the information for learning sewing stitches was the same. However the manner of presentation was varied. Therefore one condition was anthropomorphic and the other was non-anthropomorphic. Also the work is closely linked with Hartson's theory of affordances applied to user interfaces. The results suggest that facilitation of the affordances in an anthropomorphic user interface lead to statistically significant results in terms of effectiveness and user satisfaction in the sewing context. Further some violation of the affordances leads to an interface being less usable in terms of effectiveness and user satisfaction.

## I. INTRODUCTION

The research area of the main author of this paper has been aiming to discover if anthropomorphism at the user interface improves interaction in terms of effectiveness and user satisfaction. While several other researchers have also been investigating similar issues for a number of years, overall there is still a lack of clear evidence to answer the basic question of whether anthropomorphism is effective and preferred by users. In previous studies by the main author (e.g. [11, 12]), it has been suggested that a lack of facilitation of the affordances at the user interface could be a primary reason for an interface not being effective and that the presence of anthropomorphism is more secondary in nature compared to the affordances being appropriately facilitated. These arguments were based on the observations of previous experiments on various prototypes and the data collected.

While it is acknowledged that context etc can play a role, the authors would argue that if the matter was so 'simply explained', the results of researchers across the world would be more closely aligned. However, this is not the case, the results overall tend to not follow a clear pattern. In some cases anthropomorphism appears to be better and in some cases the opposite seems true.

Anthropomorphism at the user interface typically involves some part of the user interface, taking on some human quality [5]. Some examples include a synthetic character acting as an assistant or a video clip of a human [2].

Therefore, this paper presents the next important stage of the authors' work. This stage has involved developing one anthropomorphic user interface that facilitates the affordances and a contrasting (but identical in content) non-anthropomorphic user interface that deliberately violates in a subtle manner the affordances (a future stage aims to try the opposite, i.e. a non-anthropomorphic interface that facilitates the affordances and an anthropomorphic interface that violates the affordances). This differs significantly from the main author's previous work, because the previous software prototypes developed and tested were not specifically designed to facilitate/violate the affordances. However their aim was to evaluate anthropomorphic feedback. Furthermore this research is novel in that as far as we know, no one has linked the affordances with anthropomorphic feedback and more practical hands-on activities such as sewing.

The next section will present some key related literature. Following the brief literature review, the details of the experiment will be presented along with the main statistically significant results. Then the paper will conclude with a discussion in terms of the results and the affordances.

## II. KEY LITERATURE

One of the purposes of this section is to inform the reader about some of the current work carried out in connection with anthropomorphic feedback and to show that there is sometimes disagreement in the results of whether such feedback is more effective and preferred by users.

David, Lu, Kline and Cai [4] report the details of a study involving three experimental conditions in the context of a quiz about ancient history. The authors were looking into different anthropomorphic cues. These were the gender of a character and attitude and user perceptions about the character connected to quiz success. The overall results of their experiment indicated that anthropomorphic cues led to users believing the character to be less friendly, intelligent and fair. This result was linked with the male character and not with the female character. This suggests that anthropomorphic feedback (at least in some forms) may not be the best approach.

Also, Prendinger, Ma and Ishizuka [19] investigated the use of eye tracking for data collection. Prendinger et al specifically tested an animated character with gestures and voice, voice only and text. The context for this work involved showing users around an apartment via a computer monitor. Their main findings were that the character condition seemed to be better for directing 'attentional focus' to various objects on the screen. However the voice only condition fostered more attention on the part of the users towards 'reference objects' on the screen. They also observed that the text only condition induced participants to look at the text more than the character, in terms of gaze points. Finally, subjective aspects were inconclusive. Despite this study having some experimental flaws, such as having very small sample sizes, it does indicate that using an anthropomorphic entity is not necessarily better than other modes.

Furthermore, Qiu and Benbasat [20] investigated anthropomorphic agents in the context of an agent recommending products to users. They wanted to see the effects on users' 'social relationship' with an e-commerce system. In their study they observed that the anthropomorphic agent had a positive influence on users in terms of 'social presence', trust, enjoyment and future use of the 'system'. Therefore this study suggests that the anthropomorphic agent fostered positive interactions and thoughts from the part of a user.

However in an earlier study by McBreen, Anderson and Jack [15] the use of different embodied agents in different retail type domains was investigated. The retail domains they tested were cinemas, travel and banking. The agents they tested were female and male in each of the domains and formal/informal dress in each of the domains. Having conducted an experiment with this scenario, the basic results they obtained were that although participants rated the systems positively overall, no significant effects were found in participants' ratings for some applications (it must be noted however that exactly what they were rating was not clear from the paper). There were also no significant effects found for questions relating to how participants perceived the 'services' they used (e.g. likelihood of using the service in the future, convenience and ease of use etc.). Participants were also given the opportunity to select their preferred domain. This gave significant results showing the cinema domain to be the preferred one. The reason participants mainly chose the cinema domain was because of the issues of trust and errors. Overall, the cinema domain was seen as less of a problem should 'something' negative happen, e.g. it is better and more acceptable to miss a movie due to some error than say a flight to some destination. There were no significant effects for the gender and the type of agent for any of the domains tested with respect to the participants being asked if they approved of the voices used. Participants were also asked if they found the voices irritating. This gave a significant result showing that the female formal voice was preferred over the male formal voice. Also significance was found for the naturalness of the voices issue, where the female voice was considered to be more natural compared with the male voice. No differences were found for issues of agent friendliness, competence and domineering attitudes. However, there was a significant effect showing that participants had a preference for the formally dressed agent in the bank domain. The last significant effect was that participants expressed the opinion that the agents in the bank domain should be formal and that the agents in the cinema domain should be informal.

The study by McBreen et al [15] is not so clear cut as the study by Qiu and Benbasat [20]. McBreen et al found issues of lack of trust and no significance in terms of reusing the system at a future time. Therefore there are some contrasts in results with these two studies in the retail/e-commerce type context.

As stated at the outset, the main author of this work has also been working in this area for a number of years (e.g. [9, 10, 11, 12, 13, 14]) and sometimes the results are not consistent with each other.

In a previous study conducted by the main author of this paper [10] in the context of PC building instructions, an anthropomorphic character condition was tested against a non-anthropomorphic text condition. For this experiment the main results for effectiveness (based on errors) were inconclusive. However the results for subjective satisfaction tended towards a preference for the anthropomorphic interface.

In contrast, another study by Murano [9] in the specific context of English pronunciation, showed with significant results that using an anthropomorphic feedback was more effective, where the feedback helped more with the users' self-correction process in pronunciation. However in terms of user satisfaction, the results were inconclusive. The experiment had Italian participants with imperfect English pronunciation. Tasks involving pronunciation exercises were used where either an anthropomorphic (video of a human) or non-anthropomorphic (guiding text and diagram) feedback was used to assist in the correction process.

This brief review of some of the work already carried out by others and the main author of this paper, indicates that the findings do not always agree with each other across the whole range of work and is therefore worthy of further study. The last part of this literature review will discuss the concepts of affordances so as to inform the reader about the concepts and more clearly see how they were used in the experiment detailed in subsequent sections of this paper.

The concept of affordances was initially devised by Gibson [7]. Gibson was the first researcher to systematically study and propose physical affordances. As the affordances in relation to a computer user interface are different to the affordances discussed by Gibson, a detailed consideration of Gibson's theory is beyond the scope of this paper.

However affordances have been reinterpreted for application to user interfaces. Norman [17, 18] and Hartson [8] are the main sources of the reinterpretations, with more lightweight contributions from Gaver [6] and McGrenere and Ho [16] where they started to apply affordances to computer systems and to decompose affordances into different components. For brevity this paper will only briefly review Hartson's contribution as in the authors' opinion Hartson's contribution is the most substantial.

Hartson [8] identifies cognitive, physical, functional and sensory affordances. His rationale is that when doing some computer related task, the users are using cognitive, physical and sensory actions. Cognitive affordances involve 'a design feature that helps, supports, facilitates, or enables thinking and/or knowing about something' [8]. One example of this aspect concerns giving feedback to a user that is clear and precise. If one labels a button, the label should convey to the user what will happen if the button is clicked. Physical affordances are 'a design feature that helps, aids, supports, facilitates, or enables physically doing something' [8]. According to Hartson a button that can be clicked by a user is a physical object acted on by a human and its size should be large enough to elicit easy clicking. This would therefore be a physical affordance characteristic. Functional affordances concern having some purpose in relation to a physical affordance. One example is that clicking on a button should have some purpose with a goal in mind. The converse is that indiscriminately clicking somewhere on the screen is not purposeful and has no goal in mind. This idea is also mentioned in McGrenere and Ho [16]. Lastly, sensory affordances concern 'a design feature that helps, aids, supports, facilitates or enables the user in sensing (e.g. seeing, feeling, hearing) something' [8]. Sensory affordances are linked to the earlier cognitive and physical affordances as they complement one another. This means that the users need to be able to 'sense' the cognitive and physical affordances so that these affordances can help the user.

The prototype developed to deliver instruction on two sewing stitches and to also take into account the issues of facilitating and violating the affordances, was in terms of the affordances as interpreted by Hartson.

### III. THE SEWING EXPERIMENT

#### A. The Two Modes of Feedback and the Affordances

The anthropomorphic condition consisted of a video of a human describing how to perform each relevant stitch. Whilst the verbal description was taking place, the presenter also performed the actual stitch with a needle and fabric. This facilitated the affordances as described above, because the cognitive affordances were facilitated in that they supported the attempt of learning how to perform a stitch. Seeing the flow of a stitch occurring in a video aided this affordance. In this type of interface physical and functional affordances were less relevant. However the sensory affordances were also facilitated because the video amply helped the sense of seeing and hearing the instructions at the same time.

The non-anthropomorphic condition used a series of static diagrams showing the various stages of the stitches. Next to each diagram there was also some explanatory text for the user. The text was the same as the verbal description given in the anthropomorphic condition. This condition subtly violated the affordances because the series of diagrams and text did not facilitate as well the user in learning or knowing about the stitches due to the static and staged nature of displaying the steps to sew a stitch. This in turn would not have facilitated the sensory affordances because some aspects of 'flow' in performing a stitch would have been lost in this condition and therefore did not aid well the human ability of seeing or observation.

#### B. Hypotheses

The aim of the experiment was to check if anthropomorphic interface feedback would be more efficient and result in better user attitude, than non anthropomorphic interface feedback in the context of learning sewing. Also the anthropomorphic feedback was developed to facilitate the affordances, while the non-anthropomorphic feedback was developed to subtly violate some of the affordances. Testing the following hypotheses was a part of achieving the above aim.

Positive Hypothesis 1a: Participants will perform better in the easier sewing tasks after being instructed by the anthropomorphic feedback.

Null Hypothesis 1b: There will be no difference in performance for the easier sewing tasks regardless of feedback mode (anthropomorphic/non-anthropomorphic).

Positive Hypothesis 2a: Participants will perform better in the more difficult sewing tasks after being instructed by the anthropomorphic feedback.

Null Hypothesis 2b: There will be no difference in performance for the more difficult sewing tasks regardless of feedback mode (anthropomorphic/non-anthropomorphic).

Positive Hypothesis 3a: Participants will feel more positive while performing the easier sewing tasks after viewing the anthropomorphic form of instructions.

Null Hypothesis 3b: For the easier sewing tasks there will be no difference in positive perceptions on the participants' part.

Positive Hypothesis 4a: Participants will feel more positive while performing the more difficult sewing tasks after viewing the anthropomorphic form of instructions.

Null Hypothesis 4b: For the more difficult sewing tasks there will be no difference in positive perceptions on the participants' part.

Positive Hypothesis 5a: Participants will have a more positive attitude overall, using the anthropomorphic application.

Null Hypothesis 5b: There will be no difference in participants' overall positive perceptions.

#### C. Participants

There were 40 participants recruited. Participants were initially approached by the experimenter and were requested to participate in a software evaluation based on sewing. They were asked only once and not pressured to take part if they did not initially agree. This was because if they did not wish to take part in the experiment, pressuring them to do so could have negatively affected the results as the participants may have been unmotivated. Once initial agreement was granted by the participants, they were asked to complete a pre-experiment questionnaire which elicited various aspects of their past experiences.

For this experiment it was deemed to be necessary to have novices in terms of sewing skills as the prototypes developed were designed with novices in mind. Therefore the recruitment aimed to find participants with no hand or machine sewing experience and no engagement in having extensively observed others sewing in the recent past. However participants with button sewing experience were allowed as it was considered that this skill would be quite common and difficult to completely filter out and that sewing a button is quite different to sewing a particular series of stitches in a particular 'pattern' on a piece of fabric. However, in order to be sure that button sewing experience did not affect the results, the data collected during the experiment was scrutinised in terms of the averages for performance variables. The results indicated that button sewing experience and lack of such experience did not greatly affect the averages (data not included for brevity).

It was a requirement to have participants that were fluent in English. Recruitment from the university population facilitated this aspect as native English speakers would be at the appropriate level and any non-native English speakers from overseas would have a minimum English level requirement for being granted a place to study at the university.

Computer experience was also a factor that was controlled. It was deemed to be important to have participants with a comfort level in using computers, applications and basic software installation. Also all participants had a minimum of between 1 and 3 years of experience and used a computer at least 2-4 times per week.

Since the tasks involved participants actually trying to sew some stitches with a needle and fabric, it was necessary to have participants with normal or corrected eyesight, no motor control impairments and no weakness in hands. Participants with such impairments could have had difficulty in using a sewing needle.

Furthermore, the participants chosen for this experiment were right handed. Left handed participants were excluded, as pilot tests were conducted with two left handed participants to see if the prototype instructions on sewing were suitable for left handed participants as well. The participants had trouble performing the tasks. One of them mentioned that any instruction which includes hand movements in a certain direction was hard to follow for left handed people. Lastly, the age of participants ranged from 20-35 years and each participant was randomly allocated to one of the two experimental conditions (anthropomorphic or non anthropomorphic).

### D. Experiment Design

The participants were recruited randomly and were allotted one of the two applications, anthropomorphic or non-anthropomorphic. The experiment was a between users design. This design was chosen so as to avoid learning effects with the relevant sewing stitches to be used.

*1) Anthropomorphic and Non-Anthropomorphic Applications:* We developed two applications, one with anthropomorphic feedback and the other with non-anthropomorphic feedback. Both applications are Windows forms applications. Both began with the instructions on how to use the application and a briefing about the experiment itself. They then contained two sewing tasks each. Each Task is a tutorial with instructions on a type of sewing stitch followed by a test to perform the sewing stitch. Both the applications had the same two stitches instructed. The medium of instruction in the anthropomorphic feedback type application was a video of a person demonstrating the stitch. On the other hand the non-anthropomorphic application had the same instructions in text format with a graphical illustration. The instructions given in both applications for each of the stitches was exactly the same, even though the medium was a video or text. This was done to ensure that the level and accuracy of information in both types of application was exactly the same, so as not to create a bias for any one type of the application.

*2) Pilot Testing:* Before the prototypes and procedure were used in the actual experiment, some pilot testing was undertaken. Regarding the stitches used, a sewing professional was consulted regarding the suitability of the stitches and the ensuing discussion led to the use of back and chain stitch. These were chosen so that the prototype concepts could be demonstrated with contrasting stitches (i.e. one easier – back stitch, and the other more difficult – chain stitch). The instructions were taken from various websites e.g. [1, 3, 21] etc.

Furthermore, testing of the instructional content was carried out with a couple of users and these led to some improved phrasing in the instructions and diagrams initially used, for the non-anthropomorphic condition. The videos used in the anthropomorphic condition were also revised to be clearer, by altering the shooting angle. A change in the shooting angle also resulted in a better contrast with the fabric, needle and thread being used.

Once these changes had been carried out, the two conditions and experiment procedure were once again tested by two participants. The results then gave confidence that the actual experiment was ready to be carried out with real participants.

### E. Variables

The independent variables were (1) the types of feedback presenting the information (anthropomorphic and non-anthropomorphic) and (2) Type of Task (performing back stitch and chain stitch), where the values obtained from the performance data were included in the analyses (i.e. not the tasks themselves).

The dependent variables were the participants' performance in carrying out the tasks and their subjective opinions.

The dependent measures were that the performance was measured by timing how long it took to complete 5 stitches, the number of correct stitches made, the number of partially correct stitches made, the number of incorrect stitches made and the number of revisits to the instructional material presented. Each stitch (back and chain stitch) was identified as being composed of several sub-stages. Therefore if a participant only achieved one or more sub-stages, but not all required sub-stages to complete a stitch, this would be categorised as a partially correct stitch. However all sub-stages completed correctly would be categorised as a correct stitch. An incorrect stitch would be one that had no correct sub-stages completed.

The subjective opinions were measured by means of a post-experiment questionnaire. The post-experiment questionnaire was divided into three main sections. The first section contained general questions regarding the appearance and organisation of the information of the prototypes, including elicitation of the participants' feelings during the interaction. The second section contained questions relating to the tasks. These questions concentrated on the participants' feelings during the stitching tasks and on the actual information relating to the instructional material. The third section was open ended in terms of participants expressing opinions for improvement for presenting such information. Furthermore, observable attitudes noted on the observation protocol used during the experiment, were also used as a measure.

*F. Materials and Apparatus*

The materials used were:

- A laptop on which the experiment took place. The screen resolution was 1200 by 800 pixels with the highest (32 bit) colour quality. The laptop speaker volume was kept to 50% and the volume of media player control in the anthropomorphic application (windows forms application) was kept on its default value.
- Threaded needle.
- Plain cloth mounted on a frame.
- Pre loaded applications (anthropomorphic and non-anthropomorphic) on the laptop.
- Observation protocol, pre-experiment and post-experiment questionnaires.
- A meeting room on the university campus was used for conducting the experiment.

*G. Procedure and Tasks*

All participants did the experiment individually in the presence of the experimenter alone. Each participant took around 25 minutes to perform the complete experiment.

Upon arrival to the venue, each participant was allotted the same desk in the room, a comfortable chair and the same laptop was given each time. These details would ensure the participants did the experiment with maximum and equal concentration and that no bias was introduced by treating participants differently. Just after being greeted in the room by the experimenter, as a method for briefing the participants on the experiment, they were asked to read a small note on a sheet of paper just as they entered. The briefing note contained some simple details about the study, its purpose and some procedural aspects the participant would need to expect. It also stated that the participants were not personally being evaluated, but that the evaluation was of the software. The experimenter did not give any instruction out verbally. This again tried to ensure the same treatment for all participants.

Before beginning the experiment the participants were given a pre-experiment questionnaire to ensure participants had the required profile (see participants section above). The pre-experiment questionnaire elicited basic demographic information, sewing experience skills, knowledge of using computers, English proficiency and possible impairments causing potential bias, e.g. hand weakness etc.

Then the participants were allotted one of the two applications randomly. Regardless of experimental condition, participants were treated the same, and asked to proceed in the same manner. There was no difference in the two experimental conditions except in the anthropomorphic feedback condition, the application had videos for the two tutorials and in the non-anthropomorphic application the tutorials consisted of text along with graphical illustrations. Therefore the instructions were exactly the same.

The participants allocated to the anthropomorphic condition also carried out a small sound check to test the volume settings. A sample piece of sound from one of the videos was played for the sound check which simply said 'Good work you've just learned back stitch'. If required they could adjust the laptop volume to their needs if it wasn't audible enough for them. As there was no sound in the non-anthropomorphic condition, this sound check step was not required.

The next stage was to commence the experiment and run the software. Firstly the software gave the participants an introduction about how to use the application. Then some information about the experiment was presented. It consisted of explaining about the two tasks they would do.

After the introduction was completed the participant clicked 'next' to begin task 1 and reach the tutorial page. Here they were asked to begin viewing or reading the tutorial on Back stitch and told that they could go through the tutorial as many times as they liked provided they told the experimenter. They were informed it would help the experiment if they gave the true figure and it was nothing to do with their performance. The alternative of showing text for an average reading time to ensure participants read the text only once was considered. However that could have led to an anxious feeling in participants and thus possibly creating bias. This was therefore not implemented. Further, the bottom of the window prompted the participants to verbally describe their feelings to the experimenter after going through the tutorial, of which the experimenter made a note. After going through the tutorial participants could go to the next window. Here they read instructions for the test. In the instructions participants were asked to try and repeat 5 continuous stitches given in tutorial 1. The participant could choose to go back and view the tutorial again on the condition that they would have to restart the test and do all 5 stitches each time they went back to view the tutorial. Before they could start the participants were given a plain piece of cloth and a threaded needle. The experimenter recorded the time to perform the test and a count was kept of the total number of times the tutorial was viewed. Also, the experimenter made a note of participants' attitude when they were carrying out the task. After finishing task 1 participants pressed 'next' to reach the 'Break' window, where they were asked to have a short two minute break.

Task 2 began after the participants completed task 1. Task 2 was designed in exactly the same way as task 1. The task proceeded in the same way with the same rules as described above. Everything was the same except the stitch participants were shown and asked to reproduce was the more difficult Chain stitch.

During each task, participants were discreetly observed and a specially designed observation protocol was used to aid this process. The protocol was used to record the time taken for each task, the number of correct, incorrect and partially correct stitches achieved and outwardly manifested participant attitudes. The protocol was particularly useful concerning the recording of stitch accuracy. This is because it is easier to see an incorrect stitch as it is being formed, rather than at the end of the process.

After finishing both tasks the participants were asked to complete a paper based post-experiment questionnaire (see Variables section for a brief description of the content of the questionnaire) using a 5 point Likert type scale. In all cases a 5 score was the most positive score that could be allocated.

*H. Results*

The data were analysed using a multifactorial analysis of variance (MANOVA) and when significance was found, the particular issues were then subjected to post-hoc testing using in all cases either t-tests or Tukey HSD tests. This was for confirmation of significance. For brevity the post-hoc test results are not presented here. Also for brevity, only the summary data concerning significant results are presented here. The distribution summary data is presented in Appendix I below.

However, the following tables in this section present the significant results for the MANOVA analysis. While all aspects in each table have their purpose and importance, the reader's main attention is drawn to the F Ratio in bold font of each table, where the discussion will tend to concentrate on these figures. Furthermore, the abbreviation DF (Degrees of Freedom) is used in each MANOVA table. Then the Experiment Discussion section below will discuss the results in light of the issues being investigated.

For the variables task 1 - number of correct stitches, experimental condition, age range and gender, there is a significant (p < 0.01) difference. The anthropomorphic condition produced significantly more correct stitches than the non-anthropomorphic group (see table I).

TABLE I. MANOVA - TASK 1 - NO. OF CORRECT STITCHES, EXPERIMENTAL CONDITION, AGE RANGE & GENDER

| Source | DF | Sum of Squares | Mean Square | F Ratio |
|---|---|---|---|---|
| Model | 5 | 129.32008 | 25.8640 | **12.6248** |
| Error | 34 | 69.65492 | 2.0487 | Prob > F |
| C. Total | 39 | 198.97500 | | <.0001 |

For the variables Task 1 – number of incorrect stitches, experimental condition, age range and gender, there is a significant (p < 0.01) difference. The anthropomorphic condition produced significantly fewer incorrect stitches than the non-anthropomorphic group (see table II).

TABLE II. MANOVA - TASK 1 – NO. OF INCORRECT STITCHES, EXPERIMENTAL CONDITION, AGE RANGE & GENDER

| Source | DF | Sum of Squares | Mean Square | F Ratio |
|---|---|---|---|---|
| Model | 5 | 104.48059 | 20.8961 | **6.9726** |
| Error | 34 | 101.89441 | 2.9969 | Prob > F |
| C. Total | 39 | 206.37500 | | 0.0001 |

For the variables Time taken for task 2, experimental condition, age range and gender, there is a significant (p < 0.01) difference. The anthropomorphic condition completed the task significantly faster than the non-anthropomorphic condition (see table III).

TABLE III. MANOVA – TIME TAKEN FOR TASK 2, EXPERIMENTAL CONDITION, AGE RANGE & GENDER

| Source | DF | Sum of Squares | Mean Square | F Ratio |
|---|---|---|---|---|
| Model | 5 | 16.540933 | 3.30819 | **4.3322** |
| Error | 34 | 25.963264 | 0.76363 | Prob > F |
| C. Total | 39 | 42.504197 | | 0.0037 |

For the variables number of correct stitches produced for task 2, experimental condition, age range and gender, there is a significant (p < 0.01) difference. The anthropomorphic condition significantly produced more correct stitches for the more difficult task, than the non-anthropomorphic condition (see table IV).

TABLE IV. MANOVA – TASK 2 NO. CORRECT STITCHES, EXPERIMENTAL CONDITION, AGE RANGE & GENDER

| Source | DF | Sum of Squares | Mean Square | F Ratio |
|---|---|---|---|---|
| Model | 5 | 131.43425 | 26.2868 | **10.8907** |
| Error | 34 | 82.06575 | 2.4137 | Prob > F |
| C. Total | 39 | 213.50000 | | <.0001 |

For the variables number of incorrect stitches produced for task 2, experimental condition, age range and gender, there is a significant (p < 0.01) difference. The anthropomorphic condition made significantly fewer incorrect stitches for the more difficult task, than the non-anthropomorphic condition (see table V).

TABLE V. TASK 2 NO. INCORRECT STITCHES, EXPERIMENTAL CONDITION, AGE RANGE & GENDER

| Source | DF | Sum of Squares | Mean Square | F Ratio |
|---|---|---|---|---|
| Model | 5 | 47.31334 | 9.46267 | **3.6318** |
| Error | 34 | 88.58666 | 2.60549 | Prob > F |
| C. Total | 39 | 135.90000 | | 0.0097 |

For the variables number of tutorial visits for task 2, experimental condition, age range and gender, there is a significant (p < 0.01) difference. The anthropomorphic condition visited/viewed the tutorial significantly fewer times, than the non-anthropomorphic condition (see table VI).

TABLE VI. MANOVA – TASK 2- NO. TUTORIAL VISITS, EXPERIMENTAL CONDITION, AGE RANGE & GENDER

| Source | DF | Sum of Squares | Mean Square | F Ratio |
|---|---|---|---|---|
| Model | 5 | 15.223326 | 3.04467 | **3.8805** |
| Error | 34 | 26.676674 | 0.78461 | Prob > F |
| C. Total | 39 | 41.900000 | | 0.0069 |

The above summary tables, concern the performance data that was recorded as part of the experiment. Now follows the summary data concerning the subjective opinions of the participants. As stated above, these opinions were elicited by means of a post-experiment questionnaire. In all cases these responses were on a five point Likert type scale where totally positive scores were ranked as 5 and totally negative scores were ranked as 1.

For the variables of having a feeling of clarity whilst going through the tutorial for task 2, experimental condition, age range and gender, there is a significant ($p < 0.01$) difference. The participants in the anthropomorphic condition rated the tutorial with a significantly higher positive score for feelings of clarity in relation to the tutorial and the second task, than the non-anthropomorphic condition (see table VII).

TABLE VII.   MANOVA – TASK 2 CLARITY OF FEELING WHILST GOING THROUGH TUTORIAL, EXPERIMENTAL CONDITION, AGE RANGE & GENDER

| Source | DF | Sum of Squares | Mean Square | F Ratio |
|---|---|---|---|---|
| Model | 5 | 28.471534 | 5.69431 | **5.9337** |
| Error | 34 | 32.628466 | 0.95966 | Prob > F |
| C. Total | 39 | 61.100000 | | 0.0005 |

For the variables of having a feeling of satisfaction whilst going through the tutorial for task 2, experimental condition, age range and gender, there is a significant ($p < 0.01$) difference. The participants in the anthropomorphic condition rated the tutorial with a significantly higher positive score for feelings of satisfaction in relation to the tutorial and the second task, than the non-anthropomorphic condition, i.e. the participants in the anthropomorphic condition felt significantly more satisfied (see table VIII).

TABLE VIII.   MANOVA – TASK 2 – FEELINGS OF SATISFACTION WHILST GOING THROUGH TUTORIAL, EXPERIMENTAL CONDITION, AGE RANGE & GENDER

| Source | DF | Sum of Squares | Mean Square | F Ratio |
|---|---|---|---|---|
| Model | 5 | 24.728413 | 4.94568 | **4.6391** |
| Error | 34 | 36.246587 | 1.06608 | Prob > F |
| C. Total | 39 | 60.975000 | | 0.0025 |

For the variables of feeling stimulated whilst going through the tutorial for task 2, experimental condition, age range and gender, there is a significant ($p < 0.01$) difference. The participants in the anthropomorphic condition rated their feelings of being stimulated with a significantly higher positive score in relation to the tutorial and the second task, than the non-anthropomorphic condition i.e. the participants in the anthropomorphic condition felt significantly more stimulated (see table IX).

TABLE IX.   MANOVA – TASK 2 – STIMULATING FEELING WHILST GOING THROUGH TUTORIAL, EXPERIMENTAL CONDITION, AGE RANGE & GENDER

| Source | DF | Sum of Squares | Mean Square | F Ratio |
|---|---|---|---|---|
| Model | 5 | 17.183223 | 3.43664 | **5.1436** |
| Error | 34 | 22.716777 | 0.66814 | Prob > F |
| C. Total | 39 | 39.900000 | | 0.0013 |

For the variables of having a feeling of satisfaction after viewing the tutorial but before doing task 2, experimental condition, age range and gender, there is a significant ($p < 0.01$) difference. The participants in the anthropomorphic condition rated their feelings of satisfaction with a significantly higher positive score in relation to the tutorial and the second task, than the non-anthropomorphic condition, i.e. the participants in the anthropomorphic condition felt significantly more satisfied after viewing the tutorial and just prior to actually doing the task (see table X).

TABLE X.   MANOVA – TASK 2 – FEELING OF SATISFACTION AFTER VIEWING TUTORIAL BUT BEFORE DOING TASK, EXPERIMENTAL CONDITION, AGE RANGE & GENDER

| Source | DF | Sum of Squares | Mean Square | F Ratio |
|---|---|---|---|---|
| Model | 5 | 19.954554 | 3.99091 | **4.0121** |
| Error | 34 | 33.820446 | 0.99472 | Prob > F |
| C. Total | 39 | 53.775000 | | 0.0057 |

For the variables of feeling stimulated after viewing the tutorial but before doing task 2, experimental condition, age range and gender, there is a significant ($p < 0.01$) difference. The participants in the anthropomorphic condition rated their feelings of being stimulated with a significantly higher positive score in relation to the tutorial and the second task, than the non-anthropomorphic condition i.e. the participants in the anthropomorphic condition felt significantly more stimulated after viewing the tutorial and just prior to actually doing the task (see table XI).

TABLE XI.   MANOVA – TASK 2 – STIMULATING FEELING AFTER VIEWING TUTORIAL BUT BEFORE DOING TASK, EXPERIMENTAL CONDITION, AGE RANGE & GENDER

| Source | DF | Sum of Squares | Mean Square | F Ratio |
|---|---|---|---|---|
| Model | 5 | 24.067950 | 4.81359 | **4.9546** |
| Error | 34 | 33.032050 | 0.97153 | Prob > F |
| C. Total | 39 | 57.100000 | | 0.0016 |

For the variables of, a feeling of clarity whilst carrying out the task, experimental condition, age range and gender, there is a significant ($p < 0.01$) difference. The participants in the anthropomorphic condition rated their feelings of clarity whilst carrying out task 2 with a significantly higher positive score, than the non-anthropomorphic condition i.e. the participants in the anthropomorphic condition felt significantly more feelings of clarity whilst doing the task (see table XII).

TABLE XII.   MANOVA – TASK 2 – FEELING OF CLARITY DURING TASK, EXPERIMENTAL CONDITION, AGE RANGE & GENDER

| Source | DF | Sum of Squares | Mean Square | F Ratio |
|---|---|---|---|---|
| Model | 5 | 22.084087 | 4.41682 | **3.7249** |
| Error | 34 | 40.315913 | 1.18576 | Prob > F |
| C. Total | 39 | 62.400000 | | 0.0085 |

For the variables concerning a feeling of satisfaction whilst carrying out the task, experimental condition, age range and gender, there is a significant ($p < 0.01$) difference. The participants in the anthropomorphic condition rated their feelings of satisfaction whilst carrying out task 2 with a significantly higher positive score, than the non-anthropomorphic condition i.e. the participants in the anthropomorphic condition felt significantly more satisfied whilst doing the task (see table XIII).

TABLE XIII.   MANOVA – TASK 2 – SATISFYING FEELING DURING TASK, EXPERIMENTAL CONDITION, AGE RANGE & GENDER

| Source | DF | Sum of Squares | Mean Square | F Ratio |
|---|---|---|---|---|
| Model | 5 | 24.871758 | 4.97435 | **5.2683** |
| Error | 34 | 32.103242 | 0.94421 | Prob > F |
| C. Total | 39 | 56.975000 | | 0.0011 |

For the variables concerning a feeling of satisfaction for the overall learning experience, experimental condition, age range and gender, there is a significant (p < 0.01) difference. The participants in the anthropomorphic condition rated their feelings of satisfaction for the overall learning experience with a significantly higher positive score, than the non-anthropomorphic condition (see table XIV).

TABLE XIV.   MANOVA – TASK 2 – SATISFYING LEARNING EXPERIENCE, EXPERIMENTAL CONDITION, AGE RANGE & GENDER

| Source | DF | Sum of Squares | Mean Square | F Ratio |
|---|---|---|---|---|
| Model | 5 | 17.433564 | 3.48671 | **3.5213** |
| Error | 34 | 33.666436 | 0.99019 | Prob > F |
| C. Total | 39 | 51.100000 | | 0.0113 |

For the variables concerning the perceived ability to remember the stitch learned, experimental condition, age range and gender, there is a significant (p < 0.01) difference. The participants in the anthropomorphic condition had a significantly higher positive score, than the non-anthropomorphic condition, i.e. participants in the anthropomorphic condition felt they would be able to remember the stitch they had learned (see table XV).

TABLE XV.   MANOVA – TASK 2 – ABILITY TO RETAIN STITCH, EXPERIMENTAL CONDITION, AGE RANGE & GENDER

| Source | DF | Sum of Squares | Mean Square | F Ratio |
|---|---|---|---|---|
| Model | 5 | 23.035186 | 4.60704 | **3.5812** |
| Error | 34 | 43.739814 | 1.28647 | Prob > F |
| C. Total | 39 | 66.775000 | | 0.0104 |

For the variables concerning the overall experience of carrying out the two sewing tasks, experimental condition, age range and gender, there is a significant (p < 0.05) difference. The participants in the anthropomorphic condition perceived the tasks overall to be significantly easier, than the non-anthropomorphic condition (see table XVI).

TABLE XVI.   MANOVA – OVERALL EXPERIENCE OF DOING SEWING TASKS, EXPERIMENTAL CONDITION, AGE RANGE & GENDER

| Source | DF | Sum of Squares | Mean Square | F Ratio |
|---|---|---|---|---|
| Model | 5 | 14.827762 | 2.96555 | **3.3445** |
| Error | 34 | 30.147238 | 0.88668 | Prob > F |
| C. Total | 39 | 44.975000 | | 0.0146 |

For the variables concerning the quantity of on-screen information, experimental condition, age range and gender, there is a significant (p < 0.05) difference. The participants in the anthropomorphic condition perceived the quantity of on-screen information to be significantly better, than the non-anthropomorphic condition (see table XVII).

TABLE XVII.   MANOVA – ENOUGH ON-SCREEN INFORMATION, EXPERIMENTAL CONDITION, AGE RANGE & GENDER

| Source | DF | Sum of Squares | Mean Square | F Ratio |
|---|---|---|---|---|
| Model | 5 | 10.582722 | 2.11654 | **2.4800** |
| Error | 34 | 29.017278 | 0.85345 | Prob > F |
| C. Total | 39 | 39.600000 | | 0.0510 |

For the variables concerning ease of understanding of the overall task instructions, experimental condition, age range and gender, there is a significant (p < 0.01) difference. The participants in the anthropomorphic condition perceived the overall task instructions to be significantly easier to understand, than the non-anthropomorphic condition (see table XVIII).

TABLE XVIII.   MANOVA – EASY TO UNDERSTAND TASK INSTRUCTIONS, EXPERIMENTAL CONDITION, AGE RANGE & GENDER

| Source | DF | Sum of Squares | Mean Square | F Ratio |
|---|---|---|---|---|
| Model | 5 | 10.432167 | 2.08643 | **5.4809** |
| Error | 34 | 12.942833 | 0.38067 | Prob > F |
| C. Total | 39 | 23.375000 | | 0.0008 |

For the variables concerning ease of understanding of the instructions for task 1, experimental condition, age range and gender, there is a significant (p < 0.05) difference. The participants in the anthropomorphic condition perceived the task 1 instructions (tutorial content) to be significantly easier to understand, than the non-anthropomorphic condition (see table XIX).

TABLE XIX.   TUTORIAL 1 - EASILY UNDERSTOOD INSTRUCTIONS, EXPERIMENTAL CONDITION, AGE RANGE & GENDER

| Source | DF | Sum of Squares | Mean Square | F Ratio |
|---|---|---|---|---|
| Model | 5 | 7.533532 | 1.50671 | **2.8083** |
| Error | 34 | 18.241468 | 0.53651 | Prob > F |
| C. Total | 39 | 25.775000 | | 0.0315 |

For the variables concerning ease of understanding of the instructions for task 2, experimental condition, age range and gender, there is a significant (p < 0.05) difference. The participants in the anthropomorphic condition perceived task 2 instructions (tutorial content) to be significantly easier to understand, than the non-anthropomorphic condition (see table XX).

TABLE XX.   MANOVA – TUTORIAL 2 – EASILY UNDERSTOOD INSTRUCTIONS, EXPERIMENTAL CONDITION, AGE RANGE & GENDER

| Source | DF | Sum of Squares | Mean Square | F Ratio |
|---|---|---|---|---|
| Model | 5 | 17.564110 | 3.51282 | **3.3610** |
| Error | 34 | 35.535890 | 1.04517 | Prob > F |
| C. Total | 39 | 53.100000 | | 0.0142 |

*I. Experiment Discussion*

The results show a clear pattern suggesting that the anthropomorphic user interface was more effective and preferred by users. As expected the first task did not show many significant differences, because this involved a simpler stitch. However, crucially, despite the first task being simpler, the errors were significantly more in the non-anthropomorphic condition. The majority of significant results pertain to the second task which involved a more complicated stitch. Therefore the manner of presenting the instructional material had a more serious effect on the success of users and their perceptions about the interface.

The results give some confidence in accepting the stated positive hypotheses, which are reproduced below for convenience and are discussed in relation to the results presented in the previous section. Also the discussion will include the details of the affordances:

Positive Hypothesis 1a: Participants will perform better in the easier sewing tasks after being instructed by the anthropomorphic feedback. This clearly was the case as can be seen in tables 21 and 22, which showed with significance that the anthropomorphic condition incurred more correct stitches and fewer incorrect stitches. However the times taken for the first task were not significantly different.

Positive Hypothesis 2a: Participants will perform better in the more difficult sewing tasks after being instructed by the anthropomorphic feedback. This was shown to be true as can be seen in tables 23 to 26. These show that the anthropomorphic condition was significantly faster at completing the task, incurred more correct stitches, fewer incorrect stitches and fewer overall tutorial visits (which implies a more immediate/confident understanding of the stitch).

The data suggest that the facilitated affordances in the anthropomorphic condition did indeed have an effect. The cognitive affordances were facilitated by means of having a video of a human demonstrating the stitches. This better supported participants as they could see a stitch being performed in full flow, which is rather different to having to interpret a series of diagrams with accompanying text (as was the case in the non-anthropomorphic condition). Also the sensory affordances were better facilitated in the anthropomorphic condition. This is because the video of a human helped users to 'see' in full flow how a stitch should be made. This aspect of 'seeing' is very much a component of sensory affordances as discussed by Hartson [8]. While the non-anthropomorphic condition also had material for a user to 'see', it was deficient in terms of 'seeing' the flow of how to perform a stitch and therefore how to more easily go from one sub-stage to the next sub-stage so as to complete a full stitch and then a subsequent series of stitches.

Positive Hypothesis 3a: Participants will feel more positive while performing the easier sewing tasks after viewing the anthropomorphic form of instructions. This is less clear that the other hypotheses discussed in this section. However the authors cautiously accept this positive hypothesis because as can be seen in table 39, the participants in the anthropomorphic condition perceived the instructions for task 1 (i.e. the tutorial) to be significantly more easily understood, which shows a slightly more positive feeling. However not included in this paper, for brevity, the means of other non-significant results are in line with the anthropomorphic feedback being more effective and eliciting more positive user perceptions. Some examples include the number of tutorial visits being fewer for the anthropomorphic feedback and all the factors eliciting user perceptions had higher positive means for the anthropomorphic feedback (i.e. the non-anthropomorphic feedback was perceived more negatively).

Positive Hypothesis 4a: Participants will feel more positive while performing the more difficult sewing tasks after viewing the anthropomorphic form of instructions. This was also shown to be true as can be seen in tables 32 to 34. These suggest that participants in the anthropomorphic condition had more positive feelings of clarity and satisfaction. Also table 40 suggests that the instructions were significantly more easily understood under the anthropomorphic condition.

Positive Hypothesis 5a: Participants will have a more positive attitude overall, using the anthropomorphic application. The significant results also suggest this to be the case. Table 35 shows that participants felt more able to remember the stitch they had learned under the anthropomorphic condition. Also tables 36 to 38 show that positive perceptions were incurred for more general aspects, e.g. the experience, quantity of on-screen information and the task instructions being understandable. Furthermore, tables 27 to 31 suggest that participants in the anthropomorphic condition maintained a more positive attitude whilst going through the various stages of the experiment, i.e. from viewing the tutorial right through to carrying out and completing the task.

As can be clearly seen, the results for positive attitudes are strongly in favour of the anthropomorphic feedback. This is also linked with the cognitive and sensory affordances being better facilitated as discussed above. This is because users will tend to have a 'feel' (i.e. they may not consciously know why or how they 'feel' about something such as an interface) for a user interface that they perceive to have helped them well (or not). This is suggested by the results for the non-anthropomorphic condition, where it was consistently rated more negatively than the anthropomorphic feedback. Clearly the fact that overall users tended to make more errors, took longer for the tasks and viewed the tutorial more will have contributed to the more negative perceptions about the 'system'.

IV. CONCLUSIONS

The results agree with the expectation that the anthropomorphic feedback which facilitated more the affordances would have been more effective and preferred by users. The fact that the non-anthropomorphic feedback subtly violated the affordances resulted in more errors being committed and as one would expect, the users perceived that the feedback was not as good as it could be and therefore under several factors (see above) rated the non-anthropomorphic feedback significantly more negatively than the participants in the anthropomorphic condition.

When particularly the cognitive affordances and sensory affordances are not facilitated, the number of errors and the time taken to complete a task are significantly increased, particularly when the task has a degree of difficulty. The user perceptions are also more negative. They tend to feel less satisfied, less stimulated, less confident and understanding is perceived to be less. The results presented above indicate this and none of the results contradict this reasoning.

V. FURTHER WORK

As stated above, the anthropomorphic condition was deliberately developed to facilitate the affordances and the non-anthropomorphic condition deliberately and subtly violated the affordances and generally the expected results are borne out. The next stage in this work should continue to address the anthropomorphism issue and a future experiment will have an anthropomorphic condition that violates the affordances and a non-anthropomorphic condition that facilitates the affordances. This way forward should give further clear indicators that the affordances are possibly more crucial for usability than the actual anthropomorphic presence.

REFERENCES

[1] Alternative Windows (2004) Backstitch, http://www.alternative-windows.com/stitches.htm, Accessed 2011.

[2] Bengtsson, B, Burgoon, J. K, Cederberg, C, Bonito, J, and Lundeberg, M. (1999) The Impact of Anthropomorphic Interfaces on Influence, Understanding and Credibility. Proceedings of the 32nd Hawaii International Conference on System Sciences, IEEE.

[3] CoatsCrafts (2011) How to Back Stitch, http://www.coatscrafts.co.uk, Accessed 2011.

[4] David, P., Lu, T., Kline, S. and Cai, L. (2007) Social Effects of an Anthropomorphic Help Agent: Humans Versus Computers, CyberPsychology and Behaviour, 10, 3. Mary Ann Liebert Inc.

[5] De Angeli, A, Johnson, G. I. and Coventry, L. (2001) The Unfriendly User: Exploring Social Reactions to Chatterbots, Proceedings of the International Conference on Affective Human Factors Design, Asean Academic Press.

[6] Gaver, W. W. (1991) Technology Affordances, Proceedings of the ACM, CHI 91, Human Factors in Computing Systems Conference, April 27 – May 2 1991, New Orleans, Louisiana, USA, p79-84.

[7] Gibson, J. J. (1979) The Ecological Approach to Visual Perception, Houghton Mifflin Co.

[8] Hartson, H. R. (2003) Cognitive, Physical, Sensory and Functional Affordances in Interaction Design, Behaviour and Information Technology, Sept-Oct 2003, 22 (5), p.315-338.

[9] Murano, P., (2002) Effectiveness of Mapping Human-Oriented Information to Feedback From a Software Interface, Proceedings of the 24th International Conference on Information Technology Interfaces, Cavtat, Croatia, 24-27 June 2002.

[10] Murano, P., Ede, C. & Holt, P. O. (2008) Effectiveness and Preferences of Anthropomorphic User Interface Feedback in a PC Building Context and Cognitive Load, 10th International Conference on Enterprise Information Systems, Barcelona, Spain, 12-16 June 2008. (c) - INSTICC.

[11] Murano, P. & Holt, P.O. (2010) Evaluation of an Anthropomorphic User Interface in a Telephone Bidding Context and Affordances, 12th International Conference on Enterprise Information Systems, Madeira, Portugal, 8-12 June. (c) - INSTICC.

[12] Murano, P., Malik, A. & Holt, P. O. (2009) Evaluation of Anthropomorphic User Interface Feedback in an Email Client Context and Affordances, 11th International Conference on Enterprise Information Systems, Milan, Italy, 6-10 May. (c) - INSTICC.

[13] Murano, P. (2005) Why Anthropomorphic User Interface Feedback Can be Effective and Preferred by Users, 7th International Conference on Enterprise Information Systems, Miami, USA, 25-28 May 2005. (c) - INSTICC

[14] Murano, P., Gee, A. & Holt, P. O. (2007) Anthropomorphic Vs Non-Anthropomorphic User Interface Feedback for Online Hotel Bookings, 9th International Conference on Enterprise Information Systems, Funchal, Madeira, Portugal, 12-16 June 2007. (c) - INSTICC.

[15] McBreen, H, Anderson, J. and Jack, M. (2000) Evaluating 3D Embodied Conversational Agents in Contrasting VRML Retail Applications. Proceedings 4th International Conference on Autonomous Agents, p. 39-45, ACM.

[16] McGrenere, J. and Ho, W. (1991) Affordances: Clarifying and Evolving a Concept, Proceedings of Graphics Interface, May 2000, Montreal, Canada.

[17] Norman, D. A. (1999) Affordance, Conventions, and Design, Interactions, May-June 1999, p.39-42.

[18] Norman, D. A. (2002) The Design of Everyday Things, Basic Books.

[19] Prendinger, H., Ma, C. and Ishizuka, M. (2007) Eye Movements as Indices for the Utility of Life Like Interface Agents: A Pilot Study, Interacting With Computers, 19, 281-292.

[20] Qiu, L. and Benbasat, I. (2009) Evaluating Anthropomorphic Product Recommendation Agents: A Social Relationship Perspective to Designing Information Systems, Journal of Management Information Systems, 25(4), p 145-181, Sharpe Inc.

[21] Stitchclub (2011) How to Sew Chain Stitch, http://www.zincdesign.co.uk/stitchclub/Chain_stitch_final.pdf, Accessed 2011.

AUTHORS PROFILE

Dr Pietro Murano is a Computer Scientist at the University of Salford, UK. Amongst other academic and professional qualifications he holds a PhD in Computer Science. His specific research areas are in Human Computer Interaction and Usability of software systems.

Tanvi Sethi obtained an MSc in Databases and Web Based Systems in 2008 and currently works in a professional capacity in the computing/software industry.

APPENDIX I - DISTRIBUTIONS

TABLE XXI. TASK 1 - NO. OF CORRECT STITCHES

| Anthropomorphic | |
|---|---|
| Mean | 4.45 |
| Std Dev | 1.2763022 |
| Std Err Mean | 0.2853899 |
| upper 95% Mean | 5.0473278 |
| lower 95% Mean | 3.8526722 |
| N | 20 |
| **Non-Anthropomorphic** | |
| Mean | 1.1 |
| Std Dev | 1.7137217 |
| Std Err Mean | 0.3831998 |
| upper 95% Mean | 1.9020464 |
| lower 95% Mean | 0.2979536 |
| N | 20 |

TABLE XXII. TASK 1 – NO. OF INCORRECT STITCHES

| Anthropomorphic | |
|---|---|
| Mean | 0.35 |
| Std Dev | 1.1367081 |
| Std Err Mean | 0.2541757 |
| upper 95% Mean | 0.8819958 |
| lower 95% Mean | -0.181996 |
| N | 20 |
| **Non-Anthropomorphic** | |

| Anthropomorphic | |
|---|---|
| Mean | 3.4 |
| Std Dev | 2.1618705 |
| Std Err Mean | 0.4834089 |
| upper 95% Mean | 4.4117866 |
| lower 95% Mean | 2.3882134 |
| N | 20 |

TABLE XXIII. TIME TAKEN FOR TASK 2

| Anthropomorphic | |
|---|---|
| Mean | 1.6985 |
| Std Dev | 0.4186102 |
| Std Err Mean | 0.0936041 |
| upper 95% Mean | 1.8944156 |
| lower 95% Mean | 1.5025844 |
| N | 20 |
| **Non-Anthropomorphic** | |
| Mean | 2.887 |
| Std Dev | 1.1482119 |
| Std Err Mean | 0.256748 |
| upper 95% Mean | 3.4243797 |
| lower 95% Mean | 2.3496203 |
| N | 20 |

TABLE XXIV. TASK 2 – NO. OF CORRECT STITCHES

| Anthropomorphic | |
|---|---|
| Mean | 5 |
| Std Dev | 0 |
| Std Err Mean | 0 |
| upper 95% Mean | 5 |
| lower 95% Mean | 5 |
| N | 20 |
| **Non-Anthropomorphic** | |
| Mean | 1.5 |
| Std Dev | 2.1884866 |
| Std Err Mean | 0.4893605 |
| upper 95% Mean | 2.5242433 |
| lower 95% Mean | 0.4757567 |
| N | 20 |

TABLE XXV. TASK 2 – NO. OF INCORRECT STITCHES

| Anthropomorphic | |
|---|---|
| Mean | 0 |
| Std Dev | 0 |
| Std Err Mean | 0 |
| upper 95% Mean | 0 |
| lower 95% Mean | 0 |
| N | 20 |
| **Non-Anthropomorphic** | |
| Mean | 2.1 |
| Std Dev | 2.1980853 |
| Std Err Mean | 0.4915068 |
| upper 95% Mean | 3.1287356 |
| lower 95% Mean | 1.0712644 |
| N | 20 |

TABLE XXVI. TASK 2 – NO. OF TUTORIAL VISITS

| Anthropomorphic | |
|---|---|
| Mean | 1.45 |
| Std Dev | 0.6863327 |
| Std Err Mean | 0.1534687 |
| upper 95% Mean | 1.7712136 |
| lower 95% Mean | 1.1287864 |
| N | 20 |
| **Non-Anthropomorphic** | |
| Mean | 2.65 |
| Std Dev | 0.9880869 |

| Anthropomorphic | |
|---|---|
| Std Err Mean | 0.220943 |
| upper 95% Mean | 3.1124389 |
| lower 95% Mean | 2.1875611 |
| N | 20 |

TABLE XXVII. TASK 2 – CLARITY OF FEELING WHILST GOING THROUGH TUTORIAL

| Anthropomorphic | |
|---|---|
| Mean | 4.45 |
| Std Dev | 0.8255779 |
| Std Err Mean | 0.1846048 |
| upper 95% Mean | 4.8363824 |
| lower 95% Mean | 4.0636176 |
| N | 20 |
| **Non-Anthropomorphic** | |
| Mean | 2.85 |
| Std Dev | 1.0894228 |
| Std Err Mean | 0.2436024 |
| upper 95% Mean | 3.3598656 |
| lower 95% Mean | 2.3401344 |
| N | 20 |

TABLE XXVIII. TASK 2 – FEELINGS OF SATISFACTION WHILST GOING THROUGH TUTORIAL

| Anthropomorphic | |
|---|---|
| Mean | 4.4 |
| Std Dev | 0.680557 |
| Std Err Mean | 0.1521772 |
| upper 95% Mean | 4.7185105 |
| lower 95% Mean | 4.0814895 |
| N | 20 |
| **Non-Anthropomorphic** | |
| Mean | 3.15 |
| Std Dev | 1.3869694 |
| Std Err Mean | 0.3101358 |
| upper 95% Mean | 3.7991217 |
| lower 95% Mean | 2.5008783 |
| N | 20 |

TABLE XXIX. TASK 2 – STIMULATING FEELING WHILST GOING THROUGH TUTORIAL

| Anthropomorphic | |
|---|---|
| Mean | 4.5 |
| Std Dev | 0.606977 |
| Std Err Mean | 0.1357242 |
| upper 95% Mean | 4.784074 |
| lower 95% Mean | 4.215926 |
| N | 20 |
| **Non-Anthropomorphic** | |
| Mean | 3.4 |
| Std Dev | 1.0462967 |
| Std Err Mean | 0.2339591 |
| upper 95% Mean | 3.8896819 |
| lower 95% Mean | 2.9103181 |
| N | 20 |

TABLE XXX. TASK 2 – FEELING OF SATISFACTION AFTER VIEWING TUTORIAL BUT BEFORE DOING TASK

| Anthropomorphic | |
|---|---|
| Mean | 4.45 |
| Std Dev | 0.6863327 |
| Std Err Mean | 0.1534687 |
| upper 95% Mean | 4.7712136 |
| lower 95% Mean | 4.1287864 |
| N | 20 |

| **Anthropomorphic** | |
|---|---|
| **Non-Anthropomorphic** | |
| Mean | 3.2 |
| Std Dev | 1.2396944 |
| Std Err Mean | 0.2772041 |
| upper 95% Mean | 3.7801948 |
| lower 95% Mean | 2.6198052 |
| N | 20 |

TABLE XXXI. TASK 2 – STIMULATING FEELING AFTER VIEWING TUTORIAL BUT BEFORE DOING TASK

| **Anthropomorphic** | |
|---|---|
| Mean | 4.5 |
| Std Dev | 0.6882472 |
| Std Err Mean | 0.1538968 |
| upper 95% Mean | 4.8221096 |
| lower 95% Mean | 4.1778904 |
| N | 20 |
| **Non-Anthropomorphic** | |
| Mean | 3.2 |
| Std Dev | 1.2814466 |
| Std Err Mean | 0.2865402 |
| upper 95% Mean | 3.7997354 |
| lower 95% Mean | 2.6002646 |
| N | 20 |

TABLE XXXII. TASK 2 – FEELING OF CLARITY DURING TASK

| **Anthropomorphic** | |
|---|---|
| Mean | 4.4 |
| Std Dev | 0.88258 |
| Std Err Mean | 0.1973509 |
| upper 95% Mean | 4.8130601 |
| lower 95% Mean | 3.9869399 |
| N | 20 |
| **Non-Anthropomorphic** | |
| Mean | 3 |
| Std Dev | 1.213954 |
| Std Err Mean | 0.2714484 |
| upper 95% Mean | 3.5681479 |
| lower 95% Mean | 2.4318521 |
| N | 20 |

TABLE XXXIII. TASK 2 – SATISFYING FEELING DURING TASK

| **Anthropomorphic** | |
|---|---|
| Mean | 4.35 |
| Std Dev | 0.8127277 |
| Std Err Mean | 0.1817314 |
| upper 95% Mean | 4.7303683 |
| lower 95% Mean | 3.9696317 |
| N | 20 |
| **Non-Anthropomorphic** | |
| Mean | 3.2 |
| Std Dev | 1.2814466 |
| Std Err Mean | 0.2865402 |
| upper 95% Mean | 3.7997354 |
| lower 95% Mean | 2.6002646 |
| N | 20 |

TABLE XXXIV. TASK 2 – SATISFYING LEARNING EXPERIENCE

| **Anthropomorphic** | |
|---|---|
| Mean | 4.45 |
| Std Dev | 0.7591547 |
| Std Err Mean | 0.1697521 |
| upper 95% Mean | 4.8052953 |
| lower 95% Mean | 4.0947047 |
| N | 20 |

| **Anthropomorphic** | |
|---|---|
| **Non-Anthropomorphic** | |
| Mean | 3.25 |
| Std Dev | 1.1641577 |
| Std Err Mean | 0.2603136 |
| upper 95% Mean | 3.7948426 |
| lower 95% Mean | 2.7051574 |
| N | 20 |

TABLE XXXV. TASK 2 – ABILITY TO RETAIN STITCH

| **Anthropomorphic** | |
|---|---|
| Mean | 4.3 |
| Std Dev | 1.0809353 |
| Std Err Mean | 0.2417045 |
| upper 95% Mean | 4.8058933 |
| lower 95% Mean | 3.7941067 |
| N | 20 |
| **Non-Anthropomorphic** | |
| Mean | 3.05 |
| Std Dev | 1.234376 |
| Std Err Mean | 0.2760149 |
| upper 95% Mean | 3.6277058 |
| lower 95% Mean | 2.4722942 |
| N | 20 |

TABLE XXXVI. OVERALL EXPERIENCE OF DOING SEWING TASKS

| **Anthropomorphic** | |
|---|---|
| Mean | 4.55 |
| Std Dev | 0.6863327 |
| Std Err Mean | 0.1534687 |
| upper 95% Mean | 4.8712136 |
| lower 95% Mean | 4.2287864 |
| N | 20 |
| **Non-Anthropomorphic** | |
| Mean | 3.5 |
| Std Dev | 1.1470787 |
| Std Err Mean | 0.2564946 |
| upper 95% Mean | 4.0368493 |
| lower 95% Mean | 2.9631507 |
| N | 20 |

TABLE XXXVII. ENOUGH ON-SCREEN INFORMATION

| **Anthropomorphic** | |
|---|---|
| Mean | 4.55 |
| Std Dev | 0.7591547 |
| Std Err Mean | 0.1697521 |
| upper 95% Mean | 4.9052953 |
| lower 95% Mean | 4.1947047 |
| N | 20 |
| **Non-Anthropomorphic** | |
| Mean | 3.65 |
| Std Dev | 1.0399899 |
| Std Err Mean | 0.2325488 |
| upper 95% Mean | 4.1367302 |
| lower 95% Mean | 3.1632698 |
| N | 20 |

TABLE XXXVIII. EASY TO UNDERSTAND TASK INSTRUCTIONS

| **Anthropomorphic** | |
|---|---|
| Mean | 4.7 |
| Std Dev | 0.4701623 |
| Std Err Mean | 0.1051315 |
| upper 95% Mean | 4.9200428 |
| lower 95% Mean | 4.4799572 |
| N | 20 |
| **Non-Anthropomorphic** | |

TABLE XXXIX. TUTORIAL 1 – EASILY UNDERSTOOD INSTRUCTIONS

| Anthropomorphic | |
|---|---|
| Mean | 4.05 |
| Std Dev | 0.8870412 |
| Std Err Mean | 0.1983484 |
| upper 95% Mean | 4.4651481 |
| lower 95% Mean | 3.6348519 |
| N | 20 |
| **Anthropomorphic** | |
| Mean | 4.8 |
| Std Dev | 0.4103913 |
| Std Err Mean | 0.0917663 |
| upper 95% Mean | 4.9920691 |
| lower 95% Mean | 4.6079309 |
| N | 20 |
| **Non-Anthropomorphic** | |
| Mean | 4.05 |
| Std Dev | 0.9445132 |
| Std Err Mean | 0.2111996 |

TABLE XL. TUTORIAL 2 – EASILY UNDERSTOOD INSTRUCTIONS

| Anthropomorphic | |
|---|---|
| upper 95% Mean | 4.4920458 |
| lower 95% Mean | 3.6079542 |
| N | 20 |
| **Anthropomorphic** | |
| Mean | 4.45 |
| Std Dev | 0.8255779 |
| Std Err Mean | 0.1846048 |
| upper 95% Mean | 4.8363824 |
| lower 95% Mean | 4.0636176 |
| N | 20 |
| **Non-Anthropomorphic** | |
| Mean | 3.25 |
| Std Dev | 1.1641577 |
| Std Err Mean | 0.2603136 |
| upper 95% Mean | 3.7948426 |
| lower 95% Mean | 2.7051574 |
| N | 20 |